\documentclass[onecollarge]{svjour2}       
\smartqed  
\usepackage{rotating}
\usepackage{graphicx}
\usepackage{epstopdf}
\def\nuc#1#2{${}^{#1}$#2}
\newcommand{\ppc}{P-PC}                          

\begin{document}

\title{An Improved Limit on Pauli-Exclusion-Principle Forbidden Atomic Transitions}

\author{S.R.~Elliott         \and
        B.H.~LaRoque 		\and
        V.M.~Gehman			\and
         M.F.~Kidd			\and
         M.~Chen
}
\institute{S.R.~Elliott  
           \and
           B.H.~LaRoque          \and
           V.M.~Gehman           \and
           M.F.~Kidd     \at          Physics Division, Los Alamos National Laboratory, Los Alamos, NM 87545 \\
              \email{elliotts@lanl.gov}             \\
           \and
          M.~Chen \at
              Department of Physics and Life Sciences, Lawrence Livermore National Laboratory, Livermore, California 94550 \\
           \and
           V.M.~Gehman \at
           	\emph{present Address: Physics Division, Lawrence Berkeley National Laboratory,Berkeley, CA 94720}
			\and
           B.H.~LaRoque \at
             \emph{Present Address: Department of Physics, University of California, Santa Barbara, CA 93106} 
}

\date{Received: date / Accepted: date}

\maketitle

\begin{abstract}
We have examined the atomic theory behind recent constraints on the violation of the Pauli Exclusion Principle derived from experiments that look for x rays emitted from conductors while a large current is present. We also re-examine the assumptions underlying such experiments. We use the results of these studies to assess pilot measurements to develop an improved test of the Principle. We present an improved limit of $\frac{1}{2}\beta^2 < 2.6\times10^{-39}$ on the Pauli Exclusion Principle. This limit is the best to date for interactions between a system of fermions and a fermion that has not previously interacted with that given system. That is, for systems that do not obviously violate the Messiah-Greenberg symmetrization-postulate selection rule.

\end{abstract}

\maketitle
 

\section{Introduction}
\label{sec:Intro}

Pauli's original idea~\cite{Pau25} for the exclusion principle was postulated to explain patterns in the periodic table. Recently there has been interest in theories that might permit a small violation of the Pauli Exclusion Principle (PEP). The introduction of Ref.~\cite{Bel10} provides a guide to the literature with regard to both theory and experiment.

Messiah and Greenberg described a superselection rule regarding the symmetrization postulate (SP) in 1964~\cite{Mes64} by noting ``In summary, for systems with a fixed number of particles, there is a superselection rule between symmetry types which permits one to insert SP in the quantum theory in a consistent way. However the postulate does not appear as a necessary feature of the
QM description of nature." The paper by Amado and Primakoff~\cite{Ama80} used different phrasing stating ``Even if some principle permitted small mixed symmetry components in wave functions that are primarily antisymmetric, and kept them small, the symmetric world Hamiltonian would only connect mixed symmetry states to mixed symmetry states, just as it connects only antisymmetric states to antisymmetric states". This argues that electrons or nucleons in higher orbits are forbidden from transitions to lower orbits regardless of the PEP and although such studies still test other prohibited processes (e.g. electron or nucleon decay), they are not explicitly a test of the PEP. Subsequently a number of authors developed models that contained small violations of the PEP~\cite{Ign87,Gre87,Gre89,Gov89,Oku87,Oku89,Gre91,Gre00,Cho01}, but still experimental efforts must confront this constraint. The experiment of Ramberg and Snow~\cite{Ram90} pioneered a technique intended to avoid the Messiah-Greenberg superselection rule by introducing new electrons into a system. These {\em new electrons} would supposedly not have an established symmetry with respect to the electrons already contained within the system, thus avoiding the constraint. To accomplish this, Ramberg and Snow (RS) ran a high electrical current through a Cu conductor and searched for evidence of x rays emitted by a PEP-forbidden transition during a capture of an electron onto a Cu atom.

The parameter $\frac{1}{2}\beta^2$ has become commonly used to define the probability for a symmetric component of a fermion system wave function in a mixed state or the probability that when fermions form a state, it is symmetric. However, it has been shown that small violations of the symmetry principle are outside the context of quantum field theory. The paper by Greenberg~\cite{Gre00} gives a succinct summary of the theoretical situation. As a result, it is likely an over-simplification to compare this parameter as deduced from different systems directly and a wide variety of tests of the PEP are warranted. For example, the recent effort by the DAMA group~\cite{Bel99,Ber09} resulted in a strong constraint, but it is subject to the Messiah-Greenberg superselection rule.  In Table~\ref{tab:PEPLimits} we summarize the previous experimental results. These tests include looking for forbidden transitions in atomic or nuclear systems as well as looking for atoms in Pauli-forbidden states. 

In Section~\ref{sec:Assumptions}, we discuss the concept of a {\em new} fermion and the assumptions underlying the various experimental results. Next in Section~\ref{sec:Theory}, we address atomic physics issues related to the capture of electrons in PEP violating processes and the impact on the derivation of limits. In Section~\ref{sec:Experiments} we summarize our experimental activities and our results. In particular, we find that Pb offers many advantages over Cu in a RS-style experiment. We describe searches for PEP-violating capture on atoms by electrons from three different origins. Finally we finish with some discussion.

\begin{table*}

\caption{A summary of previous limits on the Pauli Exclusion Principle. $\breve{A}$ indicates an atom where the inner-most shell has 3 electrons instead of 2. $\tilde{A}$ 
indicates a nucleus with added nucleons in the ground state. The classification by Type is described in the text. $e^-_I$ refers to an electron that is part of a current, $e^-_f$ refers to an electron within the Fermi sea of a metal, and $e^-_{pp}$ refers to an electron produced by pair production.}
\label{tab:PEPLimits}

\begin{tabular}{|c|c|c|c|c|}

\hline  
Process			                              & Type   &  Experimental Limit          & $\frac{1}{2}\beta^2$ limit    & Reference \\
\hline\hline
\multicolumn{4}{|c|}{Atomic Transitions}  \\
\hline
$\beta^- + Pb \rightarrow \breve{Pb}$         &Ia&                              & $3 \times 10^{-2}$          	& \cite{Gol48}   \\
$e^-_{pp} + Ge \rightarrow \breve{Ge}$         &Ia&                              & $1.4 \times 10^{-3}$          	& This Work   \\
$e^-_I + Cu \rightarrow \breve{Cu}$             &II&                              & $1.7 \times 10^{-26}$          & \cite{Ram90}   \\
$e^-_I + Cu \rightarrow \breve{Cu}$             &II&                              & $4.5 \times 10^{-28}$          & \cite{Bar06}   \\
$e^-_I + Cu \rightarrow \breve{Cu}$             &II&                              & $6.0 \times 10^{-29}$          & \cite{Bar09}   \\
$e^-_I + Pb \rightarrow \breve{Pb}$             &II&								 & $1.5 \times 10^{-27}$          & This Work   \\
$e^-_f + Pb \rightarrow \breve{Pb}$             &IIa&								 & $2.6 \times 10^{-39}$          & This Work   \\
$I \rightarrow \breve{I} + $x ray            &III& $\tau > 2 \times 10^{27}$sec    & $3 \times 10^{-44}$            & \cite{Rei74}   \\
$I \rightarrow \breve{I} + $x ray            &III& $\tau > 4.7 \times 10^{30}$sec    & $6.5 \times 10^{-46}$            & \cite{Ber09}   \\
\hline\hline
\multicolumn{4}{|c|}{Nuclear Transitions}  \\
\hline\hline
$^{12}C \rightarrow ^{12}\tilde{C} + \gamma$  &III& $\tau > 6 \times 10^{27}$y &     $1.7 \times 10^{-44}$        & \cite{Log79}   \\
$^{12}C \rightarrow ^{12}\tilde{C} + \gamma$  &III& $\tau > 4.2 \times 10^{24}$y &                               & \cite{Arn99}   \\
$^{12}C \rightarrow ^{12}\tilde{C} + \gamma$  &III& $\tau > 5.0 \times 10^{31}$y &   $2.2 \times 10^{-57}$      & \cite{Bel10}   \\
$^{16}O \rightarrow ^{16}\tilde{O} + \gamma$  &III& $\tau > 4.6 \times 10^{26}$y &    $2.3 \times 10^{-57}$         & \cite{Suz93}   \\
$^{12}C \rightarrow ^{12}\tilde{N} + \beta^- + \bar{\nu}_e$  &IIIa& $\tau > 3.1 \times 10^{24}$y &                               & \cite{Arn99}   \\
$^{12}C \rightarrow ^{12}\tilde{N} + \beta^- + \bar{\nu}_e$  &IIIa& $\tau > 3.1 \times 10^{30}$y &                               & \cite{Bel10}   \\
$^{12}C \rightarrow ^{12}\tilde{N} + \beta^- + \bar{\nu}_e$  &IIIa& $\tau > 0.97 \times 10^{27}$sec &    $6.5 \times 10^{-34}$  & \cite{Kek90}   \\
$^{12}C \rightarrow ^{12}\tilde{B} + \beta^+ + \nu_e$  &IIIa& $\tau > 2.6 \times 10^{24}$y &                               & \cite{Arn99}   \\
$^{12}C \rightarrow ^{12}\tilde{B} + \beta^+ + \nu_e$  &IIIa& $\tau > 2.1 \times 10^{30}$y &  $2.1 \times 10^{-35}$        & \cite{Bel10}   \\
$^{12}C \rightarrow ^{11}\tilde{B} + p$    &III& $\tau > 8.9 \times 10^{29}$y &    $7.4 \times 10^{-60}$               & \cite{Bel10}   \\
$^{23}Na \rightarrow ^{22}\tilde{Ne} + p$    &III& $\tau > 7 \times 10^{24}$y &        $10^{-54}$                       & \cite{Ber97}   \\
$^{127}I \rightarrow ^{126}\tilde{Te} + p$    &III& $\tau > 9 \times 10^{24}$y &        $10^{-54}$                       & \cite{Ber97}   \\
$^{23}Na \rightarrow ^{22}\tilde{Ne} + p$    &III& $\tau > 5 \times 10^{26}$y &        $2\times10^{-55}$                       & \cite{Ber09}   \\
$^{127}I \rightarrow ^{126}\tilde{Te} + p$    &III& $\tau > 5 \times 10^{26}$y &        $2\times10^{-55}$                       & \cite{Ber09}   \\
neutron emission from Pb    					&III& $\tau > 1.0 \times 10^{20}$y &                               & \cite{Kis92}   \\
$^{12}C \rightarrow ^{11}\tilde{C} + n$    &III& $\tau > 3.4 \times 10^{30}$y &                               & \cite{Bel10}   \\
$^{16}O \rightarrow ^{15}\tilde{O} + n$    &III& $\tau > 1.0 \times 10^{20}$y &                               & \cite{Kis92}   \\
$^{16}O \rightarrow ^{15}\tilde{O} + n$    &III& $\tau > 3.7 \times 10^{26}$y &                               & \cite{Bac04}   \\
$^{12}C \rightarrow ^{8}\tilde{Be} + \alpha$    &III& $\tau > 6.1 \times 10^{23}$y &                               & \cite{Bac04}   \\
$Na/I \rightarrow \tilde{Na}/\tilde{I} \rightarrow X$ &III& $\tau > 1.7 \times 10^{25}$y & $1.5 \times 10^{-53}$   & \cite{Eji93} \\
\hline\hline
\multicolumn{4}{|c|}{Nuclear Reactions}  \\
\hline\hline
$^{12}C + p \rightarrow ^{12}\tilde{C} + p'$ &II& $\frac{d\sigma}{d\Omega}(51^o) < 40 fb/sr $   &  &  \cite{Mil90} \\
$^{12}C + p \rightarrow ^{9}\tilde{B} + \alpha$ &II& $\frac{d\sigma}{d\Omega}(51^o) < 56 fb/sr $ &  & \cite{Mil90} \\
\hline\hline
\multicolumn{4}{|c|}{Anomalous Nuclear Structures}  \\
\hline\hline 
$^5\tilde{He}/^4He$                                  &I&  $<2 \times 10^{-15}$             &                          & \cite{Nol91}  \\
$^5\tilde{Li}/^6Li$, $^6Li \rightarrow ^{6}\tilde{Be} + e^- + \bar{\nu}_e$ &III&  $<2.1 \times 10^{-15}$   &  $<5 \times 10^{-33}$      & \cite{Nol91}  \\
\hline\hline 
\multicolumn{4}{|c|}{Anomalous Atomic Structures}  \\
\hline\hline 
$\breve{Be}/Be$                                 &III&  $<9 \times 10^{-12}$             &                          & \cite{Jav00}  \\
$^{12}\breve{C}/^{12}C$                        &III&  $<2.5 \times 10^{-12}$             &                          & \cite{Bara98}  \\
$^{20}\breve{Ne}/^{20}Ne$                        &III&  $<2 \times 10^{-21} $            &                          & \cite{Nov90}  \\
$^{36}\breve{Ar}/^{36}Ar $                       &III&  $<4\times10^{-17}$              &     for H-like ions                     & \cite{Nov90}  \\
Search for $[1s2s]_{antisym}^1S_0$ state in He &III&  & $=(0.2 \pm 5.0) \times 10^{-6}$ & \cite{Dei95} \\
\hline\hline 
\multicolumn{4}{|c|}{Neutrino Statistics}  \\
\hline\hline
$^{100}Mo \rightarrow ^{100}Re + 2\beta^- + 2 \mbox{ bosionic }\bar{\nu}_e$ &Ia& $sin^2\chi < 0.6$  &               &\cite{Bara09}  \\
\hline\hline 
\multicolumn{4}{|c|}{Astrophysics and Cosmology}  \\
\hline\hline 
Solar burning and p-p bound state  &IIa      &&  $<1.6 \times 10^{-15}$     &  \cite{Pla89}  \\
Primordial nucleosynthesis and $^5\tilde{Li}$  &I& $\tilde{^5Li}/^6Li < 8 \times 10^{-18}$  & $<2 \times 10^{-28}$  & \cite{Tho92}  \\
Supernova neutrons and anomal. nuclei  &Ia& $\tilde{O}$/O $<10^{-18}$ &    $<10^{-17}$      &    \cite{Baron99} \\
Neutrino stat. and primordial nucleosyn.	&I&	$^4$He production	& 	& \cite{Dol05} \\
Thermal evolution of the Universe		&I&		&	$<10^{-15} - 10^{-17}$		& \cite{Gre89} \\
\hline\hline
\end{tabular}
\end{table*} 

\section{The {\em New} Electron Conundrum}
\label{sec:Assumptions}
When a fermion initiates an interaction with a system, it will form a total wave-function that is antisymmetric in the interchange of any two identical fermions in the system. However, if the PEP is violated, then it is possible that on some rare occasions, the resulting wave-function may be symmetric. Once the symmetry of the system is established, however, the Messiah-Greenberg superselection rule indicates that the transition probability between the two symmetry states is zero. Therefore to avoid this constraint, we must search for processes where a {\em new} fermion interacts with a system containing identical fermions and forms a symmetric state. How we define the term {\em new} in this context relates to the assumptions under which any given test of the PEP is done. The VIolation of Pauli exclusion principle (VIP) collaboration began a speculative discussion on this topic~\cite{Bar09}. Here we expand on that discussion by classifying various levels of newness and state the underlying assumptions. 
This list is ordered in decreasing confidence that the fermion-system interaction is new. We define:

\begin{itemize}
\item Type I interactions are between a system of fermions and a fermion that has not previously interacted with any other fermions.
\item Type II interactions are between a system of fermions and a fermion that has not previously interacted with that given system.
\item Type III interactions are between a system of fermions and a fermion within that given system.
\end{itemize}

In the following paragraphs we expound on these definitions in the context of Table~\ref{tab:PEPLimits}.

\begin{itemize}
\item {\bf Type-I} Primordial System Formation: Soon after the Big Bang, when the particle content of the Universe was just coming into being, all fermions would be new fermions. If some fraction of them formed symmetric states, they might have survived until the present epoch. The standard of this type of test is a search for anomalous nuclear state $^5\tilde{Li}$~\cite{Tho92}. The survival of an anomalous atomic state within a chemical environment over cosmic time scales seems less likely and we don't consider those as examples of this Type.

\item {\bf Type-Ia} Recently Created Fermions Interacting with a System: The original experiment along this line was that of Goldhaber and Scharff-Goldhaber~\cite{Gol48} where they searched for PEP violating capture of \nuc{14}{C} $\beta$ rays onto Pb atoms. The fundamental point being that the $\beta$ particles were electrons that had no previous interaction with the target atoms. That experiment was the best example of this type prior to this work, although the limit is not very restrictive. Such recently-created fermions could also be produced by pair production or nuclear reactions. 
 
\item {\bf Type-II} Distant Fermions Brought to Interact with a System: The Ramberg-Snow experiment is the classic example of this approach with the best previous limit by VIP. The current electrons through a conductor were assumed to have no previous contact with the target and therefore the experiment searched for PEP-violating transition x rays from that conductor. This assumption has some subtlety, however. The power supplies used by these experiments use AC current from modern power grids. Hence the electrons that comprise the current originate from the conduction electrons within the target sample and  circuitry that joins the sample to the power supply. The electrons are recirculated through the power supply. Therefore one might assume the fraction of these electrons that are within the target to be less {\em new} than electrons that were originally part of the other circuitry. Using a battery to produce the electrons might be more in line with this category of new fermions, however it would be difficult to sustain large current for an extended time with that technique. 
\\
\\
Corinaldesi~\cite{Cor67} suggested that the anti-symmetry of half-integral spin particles under
exchange is not a kinematic principle but  rather the time-dependent consequence
of interactions among the particles and a newly formed system may undergo PEP-violating transitions, whose probability decreases in time. Shimony~\cite{Shi06} proposed an experiment to test this hypothesis using crossed Ne ion and electron beams. Although this proposal fits our definition of a Type II experiment, the added time-dependence is a twist.

\item {\bf Type-IIa} Nearby Fermions Brought to Interact with System: The electrons in the Fermi sea of a conductor will interact with a specific atom in that conductor rarely. Because the time scale for a given electron to interact with a given atom is long, one could argue that each interaction is a legitimate new test of the PEP. The Ramberg-Snow style experiments can all be analyzed this way. Although the Fermi sea electrons and the atomic system electrons are both systems that have been established as antisymmetric, the interaction of the two can be assumed to be a new interaction.

\item {\bf Type-III} Stable System Transition: The Reines-Sobel experiment, of which the DAMA result is the best to date, defines this model as an system of electrons in an established symmetry state. A search is conducted for a spontaneous PEP-violating transition of one of the electrons. There are also many examples of experiments looking for similar processes in nuclei. All violate the Messiah-Greenberg superselection rule.

\item {\bf Type-IIIa} Stable System Transition with Particle Transmutation: The search for $\beta$ decays that can only occur if the PEP is violated is an interesting case. Such decays take place in a localized anti-symmetric nuclear system, however the nucleon created in the final PEP-violating state changes charge. This begs the question: Is this a freshly created fermion interacting with a system with which it has had no prior contact? We argue that the answer is no, because since the fermions never leave the nucleus, the PEP test is on-going within a localized system.

\end{itemize}

\section{Atomic Theory}
\label{sec:Theory}
The theory of the capture of a free electron onto an atom via a PEP violating process has not been studied well in the literature. To quantitatively describe this process requires an estimate of the probability that it will be captured ($P_{cpt}$), a description of the cascade process and transition branching ratios as the electron proceeds toward the ground state, and finally the energies of x rays emitted during this cascade. The experimental searches for PEP violation in this report all relate to the capture of electrons by an atom. In this section therefore, we examine these atomic physics issues upon which our derived limits depend.

\subsection{The Capture Probability}

When an electron collides with an atom, the probability that it will be captured ($P_{cpt}$) was assumed by RS to be greater than 10\%~\cite{Ram90}. The VIP collaboration used the same estimate so results could be compared.  $P_{cpt}$, however, likely depends on the atomic number of the target atom. To better understand the capture cross section, we considered previous calculations of muon capture and direct radiative recombination. The PEP-violating capture of an electron is analogous to the capture of a muon, if the muon mass was that of an electron. Previous estimates of the muon capture cross sections use the approximation that the muon mass is much greater than the electron mass and use classical equations of motion for the muon~\cite{Haf74}. These assumptions will not be valid for particles with mass equal to that of the electron. In contrast, direct radiative recombination (DRR) cross sections calculated for electron capture on ions holds more promise. A modified Kramer's formula~\cite{Kim83} has been shown to effectively reproduce the DRR cross section. The accuracy of this approach has been investigated and verified~\cite{Zer98} to low electron energies applicable to the Fermi sea in a metal.

To estimate $P_{cpt}$, we use the modified Kramer's formula of Ref.~\cite{Kim83} and make two assumptions. First, we assume that this formula and its expression for the effective Z is a reasonable approximation for a neutral atom. ($Z_{eff}= \frac{1}{2}(Z + Z_{ion}$), where $Z_{ion}$ is the ionization state of the atom and is equal to zero for a neutral atom.) Second, we calculate the total cross section by summing over all atomic levels (that is for all $n$), instead of only summing over open shell levels. This latter point simply states that a PEP-violating transition can be to any of the atomic shells. In our analysis, we search for 2-1 transitions, therefore we calculate the partial cross section by summing over $n \geq 2$. The cross section is then given by:

\begin{equation}
\sigma_{D} = \sum_{n\geq2}\frac{8\pi}{3\sqrt{3}}\frac{\alpha^5}{n^3}\frac{Z_{eff}^4}{K(K + E_n)}
\label{eqn:kramers}
\end{equation}

\noindent where $E_n = (13.6$ eV$)\frac{Z_{eff}^2}{n^2}$ is the binding energy of level $n$, and $K$ is the incident electron energy.

In Pb (Cu) the Fermi energy is 9.47 eV (7.0 eV)~\cite{CRC91}, and Eqn.~\ref{eqn:kramers} gives $1.2 \times 10^{-18}cm^2$ ($1.9 \times 10^{-19} cm^2$). One can compare this to the cross section ($\sigma_e$) for an interaction between a conduction electron and an atom. The mean free path ($\mu$) for an current electron  in Pb (Cu) is 2.34$\times10^{-7}$ cm (3.91$\times10^{-6}$ cm) and is determined by the resistivity of the metal and its Fermi energy. Using values of $\mu$ and the atomic density in the metal from Ref.~\cite{CRC91}, a cross section can be estimated for Pb (Cu) as $\sigma_e = 1.3\times10^{-16}$ cm$^2$ ($3.0\times10^{-18}$ cm$^2$). The ratio ($\sigma_{D}/\sigma_e$) of these two cross sections is an estimate of $P_{cpt}$ with the result that $P_{cpt}$ = 0.009 (0.058) for Pb (Cu).

\subsection{The Cascade}
When an electron is captured, it cascades through the energy levels eventually emitting a K$_{\alpha}$ x ray as it reaches the ground state. Although higher order transitions such as K$_{\beta}$ are possible, it is estimated that these transitions would have a reduced intensity as is seen in muonic x rays and in x ray emission during electron capture on ions. In the VIP analysis, the Cu K$_{\alpha}$ line is not resolved into the K$_{\alpha 1}$ and K$_{\alpha 2}$ components and the corresponding forbidden lines blend into a lone peak 300 eV lower in energy (see Table~\ref{tab:PEPTheory}). However, the VIP analysis did not correct for the possible emission of a forbidden x ray that is the analog of the K$_{\beta}$. This would be a modest correction to their efficiency.  

\subsection{The X-Ray Energies}
If a {\em new electron} makes a Pauli-forbidden transition in an atom, one would expect an x-ray emission similar to the K$_{\alpha}$ transition in the host material. However during this process the K shell contains 2 electrons, unlike a commonplace K$_{\alpha}$ transition, and therefore the energy is shifted down somewhat due to the additional shielding of the nuclear charge. The energies of these transitions were calculated with an estimated accuracy of a few eV and are given for a few key elements in Table~\ref{tab:PEPTheory}. These results are based on the Dirac-Hartree-Slater model with Breit interaction and QED corrections. These are relativistic jj configuration average calculations that include relaxation effects by performing separate self-consistent field calculations for initial and final states. The algorithm used to calculate these transition energies was modified to allow 3 electrons in 1s shell. (See Ref.~\cite{Bar06} for an independent estimate of the size of this shift for Cu atoms for which the calculation by one of us (M.C.) gives a similar result.) The estimated energy of the Pauli forbidden K$_{\alpha 2}$ transition in Pb is then 71.6 keV and would appear just below the normal 72.8-keV Pb x ray.   

\begin{table}
\caption{The atomic transitions resulting from violation of the Pauli Exclusion Principle, indicated by the column labeled {\em forb.}. For reference, the allowed transition energies are also quoted ({\em allow.}). Energies are in eV.}
\label{tab:PEPTheory}
\begin{tabular}{|c||c|c||c|c||c|c||}
\hline\hline
Transition 						&	\multicolumn{2}{|c||}{Cu}		& 	\multicolumn{2}{|c||}{Ge}		& \multicolumn{2}{|c||}{Pb}		\\
\hline							&	forb.	&	allow.			&	forb.		&	allow.		&	forb.		&	allow.		\\
\hline
1s - 2p$_{3/2}$ K$_{\alpha 1}$	&	7741		&	8047				&  9543		&	9886				&	73713		&	74961	 \\
1s - 2p$_{1/2}$ K$_{\alpha 2}$	&	7723		&	8027				&	9516		&	9854				&	71652		&	72798	 \\
2p$_{3/2}$ - 3s					&	738		&					&	953			&		 		&	8920			&			\\
2p$_{1/2}$ - 3d$_{3/2}$			&873			&	951				&	1131			&1221			&12241			&	12611	 \\
2p$_{3/2}$ - 3d$_{3/2}$			&856			&	931				&	1104			&1189			&	10180		&	10448	 \\
2p$_{1/2}$ - 3s					&	755		&					&	981			&				&	10981		&			 \\
2p$_{3/2}$ - 3d$_{5/2}$			&856			&	931				&	1104			&	1190			&	10276		&	10550	 \\
\hline\hline
\end{tabular}
\end{table}

\section{The Experiments}
\label{sec:Experiments}

 In this work we investigate improving upon the RS technique by using Pb instead of Cu as the conductor. Pb has a higher resistivity which leads to more electron-atom collisions. It also produces higher-energy x rays, which are less attenuated by self-shielding, and populate spectra in a region of lower relative background. Finally, the increased separation between the K$_{\alpha}$ emission from Pb and the PEP forbidden transition also results in a lower background under the searched-for peak. However, a result of the use of Pb is that the various possible PEP forbidden transitions are well separated in energy and do not blend. Hence the efficiency must be considered for each specific transition within the search.
 
The recent use of p-type, point-contact Ge detectors (PPC) for dark matter and double beta decay searches~\cite{Aal10} provide an opportunity for PEP-forbidden transition studies. These detectors have a much lower capacitance than the more commonly used semi-coax design and hence have excellent resolution at low energies even in sizable detectors. As a result, line features due to x-ray emission  are well resolved. This permits a search for x-ray emission due to PEP-forbidden transitions. In our experiment we use a PPC built by ORTEC~\cite{ORTEC} as a prototype detector for the {\sc Majorana Demonstrator}~\cite{ell08,gui08}. This detector is 53.7 mm long and 66.5 mm in diameter. It has a 3-mm diameter contact and a bevel on the edge of the contact end that is a 6 mm by 6 mm right triangle. The dead layer thickness ($0.97\pm0.03$ mm) was determined by source studies similar to those described in Ref~\cite{Bud09}. 

In this section, we consider two searches for PEP-violating capture on Pb and one on Ge. The first of the Pb experiments is a RS-style, Type II experiment using Pb instead of Cu. The second is an analysis of the same data but considering all the free electrons in the conductor as the interacting fermions. This is a Type IIa experiment by our nomenclature. Finally, we look at electrons from pair production capturing on Ge atoms; a Type Ia experiment. Each of these three searches is described in turn in subsections below.

\subsection{Current Through Lead}
The result of the VIP experiment is based on the RS concept and provides the best previous limit on Type I or II experiments. The VIP effort improved on the RS limit by using higher currents and lower background~\cite{Bar09}.

\begin{figure}
\includegraphics[width=8cm]{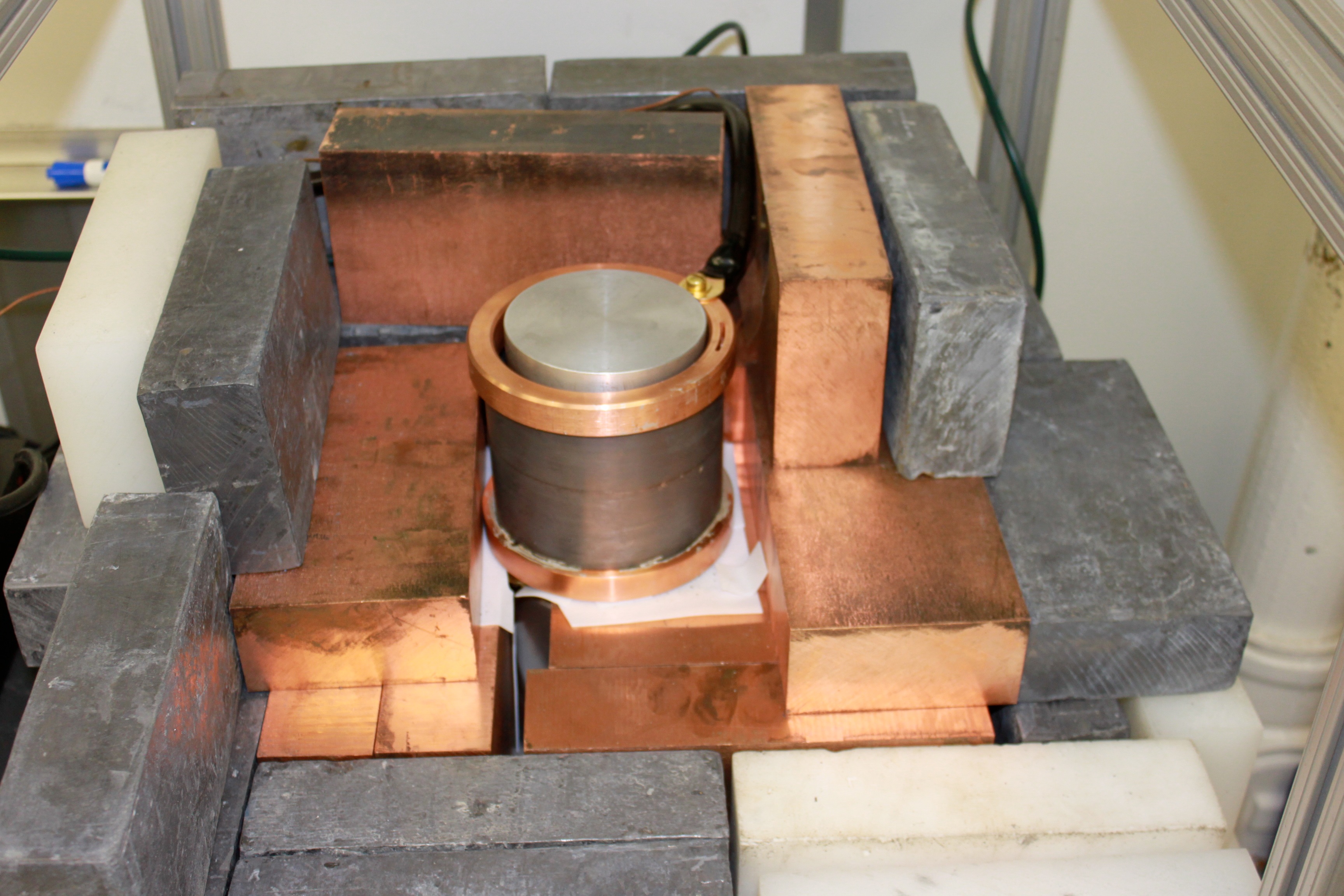}
\caption{A photograph of the experimental setup with much of the shielding removed. The Pb conductor with its Cu contacts is shown surrounding the Ge detector.}
\label{fig:Setup}
\end{figure}

In our work, a Pb cylinder 1.15 mm thick with an inner diameter of 11.25 cm surrounds the detector. The length of the Pb cylinder (D) is 8.89 cm. The ends of this Pb cylinder are attached to Cu rings with conductive epoxy and these rings provide electrical contact to the Pb. Figure~\ref{fig:Setup} shows the key aspects of our experimental setup. The detector was surrounded by 5 cm of Cu and 5 cm of Pb as a shield. The experiment was conducted in a basement laboratory at 2260 m with minimal overburden.

\begin{figure}
\includegraphics[width=10cm]{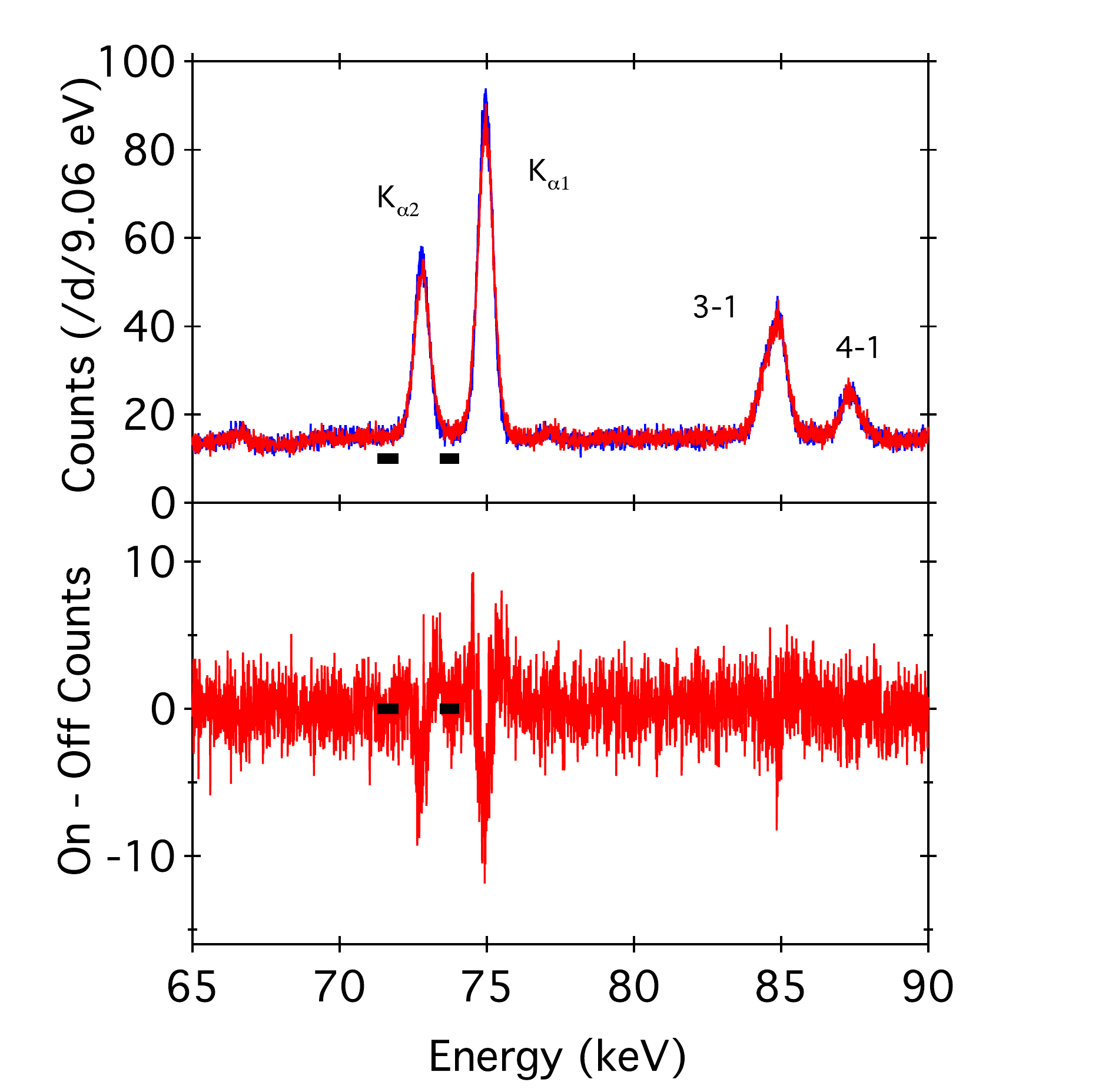}
\caption{The spectra of the data observed with the current on and off (top) and a difference spectrum between data taken with the current on and that taken with the current off (bottom). The two spectra in the top panel are very similar. The thick bar indicates the forbidden-transition region of interest. The four peaks are x rays originating from fluorescence of the Pb.}
\label{fig:Data}
\end{figure}

The current through the Pb conductor was 110 A at $\approx$0.5 V. The system was current controlled and the actual voltage varied a few per cent with temperature. We collected 254 (258) hours of current-on (current-off) data. The spectra were acquired using ORTEC NIM electronics read out using the ORCA data acquisition software~\cite{How04} and are shown in Fig.~\ref{fig:Data}. The FWHM of the lines in this energy region is about 1.15 keV for the detector used in this work.
The detector resolution at low energies is moderately sensitive to electronic noise. The width of the Pb x-ray peaks increased by 7\% when the current was on. This effect is also seen in the noise wall at low energies, which increases from 750 eV to 1000 eV with the current on. This small change in the spectrum does affect our analysis of the PEP forbidden peak as it increases the background in the region of interest and weakens the deduced constraint on $\frac{1}{2}\beta^2$. It also explains the structure in the difference spectrum of Fig.~\ref{fig:Data}.

For easier comparison of our Pb data to the VIP Cu data, we sum the K$_{\alpha 1}$ and K$_{\alpha 2}$ lines, but exclude K$_{\beta}$ and accept the additional efficiency penalty.
 To estimate the fraction of captured electrons that emit this K$_{\alpha 2}$ x ray, we assume that the forbidden emission spectrum mirrors that of the allowed emission.
 In the data, $\sim$70\% of the Pb x rays are K$_{\alpha}$ with the remaining being 3-1 or 4-1 transitions. (See Fig.~\ref{fig:Data}.)

Because the two PEP violating x rays are near the Pb K$_{\alpha}$ lines, we chose our regions of interest to minimize the ratio, $\sqrt{B}/\epsilon_{ROI}$, where $B$ is the background within the window and $\epsilon_{ROI}$ is the efficiency factor due to the fraction of the line shape contained within the region of interest. In a flat background spectrum, a symmetric window of width 2.8$\sigma$ is the optimum region-of-interest choice. For our data we calculate the optimum window taking into account the location and width of the neighboring peaks, whose tails can increase $B$ if the region of interest is chosen too wide. The deduced region of interest  (see Fig.~\ref{fig:Data}) for K$_{\alpha 1}$ is 73.395 - 74.066 keV with an efficiency of 0.811 and for K$_{\alpha 2}$ is 71.295 - 72.002 keV with an efficiency of 0.834. The relative probability of these two transitions is estimated to be $\epsilon_{BR}$ = 0.47 and 0.23 respectively. The efficiency for detecting x rays from the Pb ($\epsilon_x$) was determined by simulation using the MaGe~\cite{Bos11} framework developed by the {\sc Majorana} and GERDA collaborations. The simulation was verified by comparing to source measurements using $\gamma$ rays from \nuc{241}{Am} and \nuc{133}{Ba} placed both inside and outside the Pb sheet. The simulation matched the measurements to 10\% at all points close to the ROI. For 71.6 keV x rays emitted uniformly from the Pb cylinder $\epsilon_x$ was determined to be 0.0072. The contributions to the x ray efficiency are summarized in Table~\ref{tab:Eff}.

\begin{table}
\caption{A summary of the contributions to the x-ray detection efficiency. $\epsilon_{tot}$ is the product of the 3 efficiency factors given in the table.}
\label{tab:Eff}
\begin{tabular}{|c|c|c|c|c|c|}
\hline\hline
Line 			&	Energy	&$\epsilon_{x}$		& $\epsilon_{ROI}$		& $\epsilon_{BR}$		& $\epsilon_{tot}$  			\\
\hline
K$_{\alpha 2}$	& 71.6 KeV	&0.0072				&	0.834				&	0.23					& 0.0014	 \\
K$_{\alpha 1}$	& 73.7 keV	& 0.0072				&	0.811				& 	0.47					& 0.0027	 \\
\hline\hline
\end{tabular}
\end{table}

\begin{table*}
\caption{A summary of the detection rates in the two peaks. Columns 4 and 5 give the rates per hour with the current on ($R_{on}$) and off ($R_{off}$) respectively. The differential rates ($\delta R$) given in the final column include the x-ray detection efficiencies. Note that an excess is found between the current on and current off configurations. This excess is due to electronic noise originating from the power supply and is discussed in the text.}
\label{tab:Rates}
\begin{tabular}{|c|c|c|c|c|c|c|}
\hline\hline
				&	Energy	&Counts On 		&Counts Off			& $R_{on}$ (/h)		& $R_{off}$	(/h)		& ($\delta R$)  (/h)			\\
\hline
K$_{\alpha 2}$	& 71.6 keV	&12503			&	12858 			& 49.22$\pm$ 0.44	& 	49.84$\pm$0.44	& -443$\pm$448	 \\
K$_{\alpha 1}$	& 73.7 keV	&12995 			&	12539 			& 51.16$\pm$0.45		&48.60$\pm$0.43		& 935$\pm$228	 \\
\hline
weighted Average &	&	&	&		&& 652$\pm$203	\\
\hline\hline
\end{tabular}
\end{table*}

 The number of counts observed with the current on (off) are given in Table~\ref{tab:Rates}. The differential rates ($\delta R$) are given by ($R_{on}$-$R_{off}$)/$\epsilon_{tot}$. We then use the weighted average of the $\delta R$ results for the two lines in the determination of upper limit on the number of events that could be due to PEP violation. The previous work of RS and VIP used a 3-$\sigma$ upper limit for the number of excess counts with the current on and for direct comparison we do the same here. It is clear that the positive excess shown in Table~\ref{tab:Rates} is due to the electronic noise with the current on and is not a PEP-violation effect. Therefore, we emphasize here that we base our limit on the 3$\sigma$ variation from the excess to be as conservative as possible. Explicitly we calculate the upper limit based on $N_{3\sigma} = ((652 + 3 \times 203) /h)(254 h)(\epsilon_x) = 2307$. We have incorporated the difference in live time in the subtraction. We include the factor $\epsilon_x$ in this expression so $N_{3\sigma}$ can be compared directly to VIP and RS. As a result, this factor also appears explicitly in Eqn.~\ref{eqn:beta}.

The number of new electrons introduced into the metal is given by $N_{new} = (1/e)\sum I\Delta t $, where $e$ is the electron charge, $I$ is the current passed through the conductor during time $\Delta t$ and the sum is over all measurement periods.
The number of interactions ($N_{int}$) by an individual electron transversing the metal is given by $\frac{D}{\mu}$, where $D$ is the distance through the conductor the electrons travel. For Pb, $N_{int}$= 3$\times10^7$.
The 3-sigma upper limit on $\frac{1}{2}\beta^2$ from our work is then given by:

\begin{equation}
\label{eqn:beta}
\frac{1}{2}\beta^2 < \frac{N_{3\sigma}}{N_{new}\epsilon_x P_{cpt}N_{int}} = 1.5 \times 10^{-27}.
\end{equation}

Table~\ref{tab:Param} summarizes the parameters for the 3 experiments of the RS genre. The reference for the latest VIP result~\cite{Bar09} does not provide full detail for their preliminary results of the underground work as was done for their above ground studies. The underground runs include nearly a half year of current-on/off data.

\begin{table*}
\caption{A comparison of the experimental parameters. The values for the VIP above ground (AG) work and underground (UG) work are quoted separately. The value for $N_{3\sigma}$ was not given by VIP (AG) and the value given is our estimate based on their limit for $\frac{1}{2}\beta^2$. In the final column, the limits on $\frac{1}{2}\beta^2$ include our estimates of the new values of the Ramberg-Snow and VIP limits based on results of our work. Specifically, we used our values of $P_{cpt}$ for these results.}
\label{tab:Param}
\begin{tabular}{|c|c|c|c|c|c|c|}
\hline\hline
Project 			&	$N_{new}$		& $\epsilon_x$		& $P_{cpt}$			& $N_{int}$  		& $N_{3\sigma}$ & $\frac{1}{2}\beta^2$	\\
\hline
RS				& 9.7$\times10^{25}$	&	0.0029			& 	0.058			& 6.4$\times10^5$	& 300			& $<2.9\times10^{-26}$\\
VIP	(AG)			& 2.2$\times10^{26}$	&	0.01				&	0.058			& 2.3$\times10^6$	& 219			& $<7.7\times10^{-28}$\\
VIP	(UG)			& 3.5$\times10^{27}$ &	0.01				&	0.058			& 2.3$\times10^6$	& $\sim$500		&$<1.1\times10^{-28}$\\
This Work		& 6.29$\times10^{26}$&	0.0072			&	0.009			& 3.8$\times10^7$	& 2307 			& $<1.5\times10^{-27}$\\
\hline\hline
\end{tabular}
\end{table*}

\subsection{Free Electrons in a Metal}
The current through a conductor in a Ramberg-Snow style experiment is comprised of electrons from the circuit itself. Since the electrons originate from the conductor, one should consider whether the current is necessary. One aspect of metal conductors is that there are a large number of free electrons unlike insulators such as NaI or semi-conductors like Ge. The interaction of these free electrons with atoms in the metal can avoid the Messiah-Greenberg superselection rule under a specific set of assumptions.

A specific free electron in the metal interacts with a specific atom very rarely. The time frame is long enough that one might assume each such interaction is a new possibility to test the PEP. That is, the electron-atom system does not remember their previous interaction. For a given electron, the time between interactions is $1.3\times10^{-15}$ s (2.5$\times10^{-14}$ s) for Pb (Cu). With approximately an Avogadro's number of atoms in a sample, the time between collisions between a given electron and a given atom is tens to hundreds of years. 

If one analyzes the data ignoring the current and instead considers free electron collisions, a much improved constraint on PEP violation is found. The expression from Eqn.~\ref{eqn:beta} can now be written,

\begin{equation}
\label{eqn:free}
\frac{1}{2}\beta^2 < \frac{N_{3\sigma}}{\epsilon_{tot}} \frac{1}{P_{cpt}N_{new}^{free}N_{int}^{free}}
\end{equation}

\noindent where $N_{int}^{free}$ and $N_{new}^{free}$ are given by

\begin{eqnarray}
N_{int}^{free} &= &  \Delta t \frac{v_f}{\mu}    \\
N_{new}^{free} &= &N_e V		\nonumber
\end{eqnarray}

\noindent where $N_e$ is the free electron density, $V$ is the volume of the sample and $v_f$ is the Fermi velocity of electrons in the metal. The factor $\frac{\mu}{v_f}$ is the time between electron-atom collisions.

In Table~\ref{tab:free} we have calculated such a limit from our data using the sum of the current off and current on spectra. We added the two spectra, found the total number of counts in the two windows and used the square root of the number of counts as estimate of the 1-$\sigma$ uncertainty.

\begin{table*}
\caption{Free electron analysis of the violation of PEP. As above, we estimated the limit on the number of x-rays detected in VIP-UG to be 500 based on their result. The paper itself did not provide that number directly.}
\label{tab:free}
\begin{tabular}{|c|c|c|c|c|c|c|}
\hline\hline
Experiment		&	$N_e$ (/cm$^3$)		& $V$ (cm$^3$)	& $v_f$	(cm/s)		& $N_{int}^{free}\times N_{new}^{free}$ & $\frac{N_{3\sigma}}{\epsilon_{tot}}$ 	& $\frac{1}{2}\beta^2$		\\
\hline
VIP-UG			& $8.41\times10^{22}$	& 1.2			& $1.57\times10^8$	&$1.03\times10^{44}$ & $5 \times10^4$						& $8.4\times10^{-39}$		\\
This Work		& $1.33\times10^{23}$	& 36.1			& $1.83\times10^8$	&$6.88\times10^{45}$ & $1.64\times10^5$					& $2.6\times10^{-39}$	\\
\hline\hline
\end{tabular}
\end{table*}

The results from Table~\ref{tab:free} are very much more restrictive than for the Ramberg-Snow approach. Although the VIP-UG experiment has a lower background and a much longer run time, our Pb sample has a much larger volume and the time between collisions is much shorter resulting in a more restrictive limit.

\subsection{Electrons from Pair Production}
If a $\gamma$ ray reacts by pair production in the Ge detector, the electron produced is new to any atomic system and may violate the PEP as it slows down. If so it will capture and cascade to the K shell emitting 10.6 keV of energy that will sum with the initial energy deposit. Hence one can search for echos to the double escape (DEP) and single escape (SEP) peaks in the spectrum. 

\begin{figure}
\includegraphics[width=6cm]{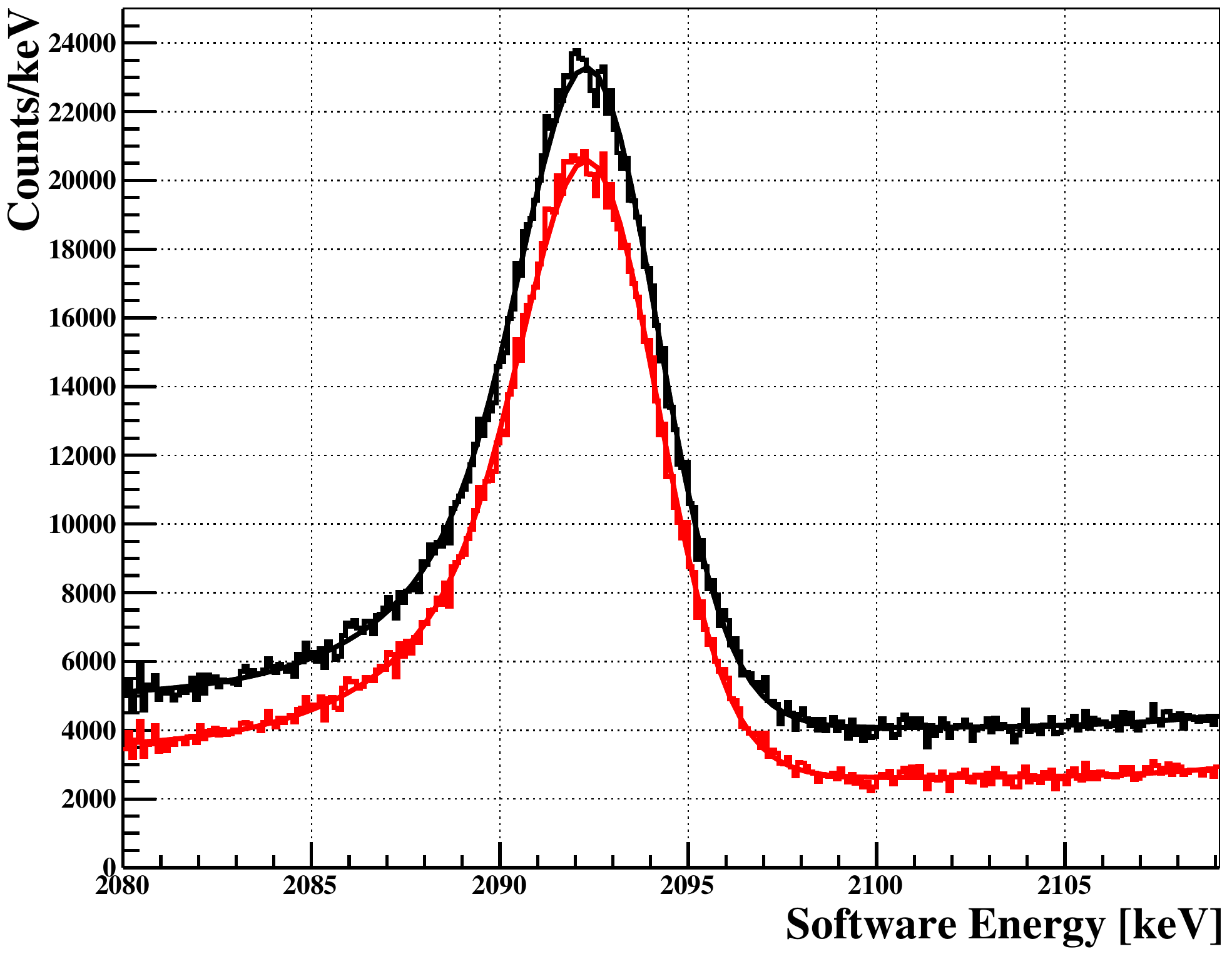}
\includegraphics[width=6cm]{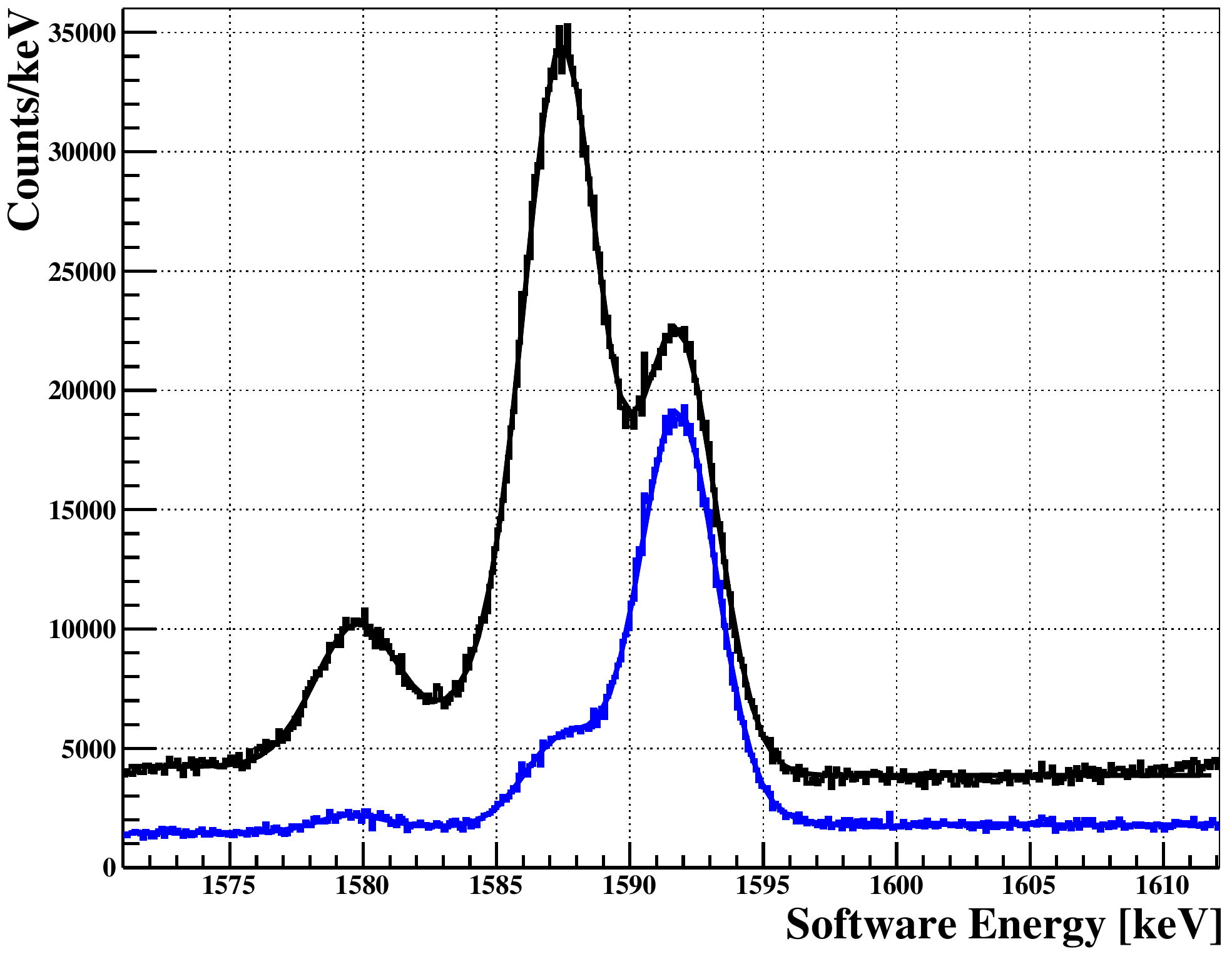}
\caption{The left panel shows a spectrum histogram near the single escape peak resulting from exposing our detector to a Th source. The higher curve is the whole data set and the lower curve shows the spectrum after implementing a waveform analysis to eliminate single site energy deposits within the crystal. The line is a fit through the data to determine peak strength. The right panel shows a similar plot of the region near the double escape peak, however the lower curve now excludes multiple site energy deposits.}
\label{fig:PhantomPeak}
\end{figure}

We exposed the Ge detector described above for 3 weeks to a Th source and then searched for the peak echos related to the DEP and SEP from the 2.6-MeV $\gamma$ ray from \nuc{208}{Tl}. For the DEP, both of the annihilation $\gamma$ rays escape the detector and the PEP-violating low-energy x-ray emissions are part of the single site energy deposit (SSE). In contrast, the SEP is by its nature a multiple site energy deposit (MSE). Though counts in the full energy peak (FEP) can result from pair production events in which the annihilation gamma rays do not escape, most arise from multiple Compton scatters.  Therefore, we do not consider the FEP in this analysis. So for each the DEP and SEP energy regions we search for a peak that is 10.6 keV above the pair production features. The spectrum near each of these peaks is shown in Fig.~\ref{fig:PhantomPeak} and in neither case is there any indication of any peak echo. In this figure, the SEP (DEP) region of the spectrum is shown before any analysis cuts and after a cut to select for MSE (SSE) deposit events. Because the low-energy x rays are emitted internal to the crystal the efficiency is effectively 100\% and the ratio of the events in any peak echo to the pair production peak is a measure of the violation of PEP. Any uncertainty in the efficiency or deviation from 100\% would cancel in the ratio. 

All events were digitized and we could analyze the event waveforms to select out SSE and MSE events using an analysis similar to Ref.~\cite{Bud09a}.
In neither case was any evidence seen and we estimated the upper limit on the existence of such a peak to be the square root of the number of counts in a window that has a width defined by the primary peak width. Table~\ref{tab:PP} summarizes the results of this Type Ia experiment.  

\begin{table}
\caption{A summary of the limits from the DEP-SSE and SEP-MSE analyses.}
\label{tab:PP}
\begin{tabular}{|c|c|c|c|}
\hline\hline
				&	Peak Counts 		&Echo Peak 			& $\frac{1}{2}\beta^2$			\\
				&					&Region Counts		&	(3 $\sigma$	limit)						\\
\hline
DEP-SSE			&	656999			& 92100				&	$ <0.0014$					\\
SEP-MSE			&	957525			& 236160				&	$ <0.0015$	\\
\hline\hline
\end{tabular}
\end{table}

\section{Discussion}
These exploratory experiments, conducted above ground with a detector inside a commercial cryostat, were able to produce competitive limits in $\frac{1}{2}\beta^2$ and one result that is the best current limit for our defined Type II interactions. The key advantages of our apparatus being the use of Pb and a \ppc\ Ge detector. There are a few obvious improvements that would enhance our work further. First, a helical wire of Pb instead of a cylinder could increase $D$ significantly. Doing this would increase the resistance and hence might require a decrease in the current. Even so, this could lead to an increase in sensitivity. Second, switching to a low-background detector operated underground would improve the sensitivity significantly. Eliminating the electronic noise pick-up associated with the current would also improve the background. And finally a longer run time would be in order. Using a \nuc{228}{Th} or \nuc{232}{U} source would improve the peak-to-continuum ratio in our Th source data.

The results using the free electron-atom collision rate in a conductor greatly improves the limit obtained on the PEP. This is due to the much larger number of individual tests of the PEP. The theoretical situation describing violations of PEP is still not entirely clear. Therefore, it is important to be clear about the assumptions that underly a test of the PEP. We have addressed this issue by categorizing the various experiments by Type defined as how {\em new} the fermion-system interaction can be assumed to be. We have also addressed a number of atomic physics issues related to these experiments.

\begin{acknowledgements}
 We gratefully acknowledge the support of the U.S. Department of Energy, Office of Nuclear Physics under
Contract No. 2011LANLE9BW. MHCÕs work was performed under the auspices of the U.S. Department of energy by Lawrence Livermore National Laboratory under Contract No. DE-AC52-07NA27344. We thank Keith Rielage and Yuri Efremenko for a careful reading of the manuscript. We thank Larry Rodriguez and Harry Salazar for helpful technical discussions.  We thank P. Vogel, R. Mohapatra, and O.W. Greenberg for useful discussions of the theory.
\end{acknowledgements}

\bibliographystyle{spmpsci}      
\bibliography{PauliExclusion.bbl}

\end{document}